\documentstyle[prl,aps]{revtex}

\begin{document}

\draft

\title{Decoherence in quantum open systems revisited}

\author{Robert Alicki}

\address{Institute of Theoretical Physics and Astrophysics, University
of Gda\'nsk, Wita Stwosza 57, PL 80-952 Gda\'nsk, Poland}

\date{\today}
\maketitle

\begin{abstract}
The following statements belonging to the folklore of the theory of environmental
decoherence are shown to be incorrect:
1) linear coupling to harmonic oscillator bath is a universal model
of decoherence, 2) chaotic environments are more efficient decoherers. 

\end{abstract}


Decoherence became one of the most popular topics in the physical literature of the
last decade [1-3]. This is mainly due to the progress in experimental techniques
allowing to observe the onset of decoherence at the most interesting regime i.e.
at the border between quantum and classical worlds [4]. Another motivation is 
a destructive role
of decoherence in the possible future technology based on
quantum information processing [5]. Despite the fact that the theoretical models of 
decoherence exist at least for 40 years [6] a closer look at certain aspects of these theories 
reveals quite fundamental inconsistencies and misconceptions. 
\par
We begin with a rather eclectic definition. 
\par 
{\it Decoherence is the irreversible, uncontrollable and persistent formation of
quantum correlations (entanglement) of the system with environment.}
\par
Usually, decoherence is accompanied by {\it dissipation} i.e. the exchange of energy
with environment. For the sake of clarity we shall restrict ourselves to the case
of {\it pure decoherence} called also {\it dephasing} for which the process of energy
dissipation is neglible. Pure decoherence is supposed to be the main ingredient of the theory
explaining the apparent absence of superpositions of macroscopically distinguishable states
and the transition from quantum to classical world. Indeed, superpositions of
quantum states separated by large energy gaps are practically not observable due to 
rapidly oscillating phases, hence the mechanism of environmental decoherence becomes 
interesting and relevant for the states with almost equal energies. 
\par
We shall concentrate ourselves on a carefull analysis of the following issues:
\par
a) No-go theorem for pure decoherence which states that this phenomenon cannot
be described  by a coupling to harmonic oscillator bath which is linear in the oscillator
coordinates and/or momenta or equivalently by a coupling to a free bose system linear
in field operators.
\par
b) False decoherence.
\par
c) Dependence of pure decoherence rate on chaotic properties of environment.
\par
The "physical proof" of the formulated above {\it no-go theorem} is very simple.
Pure decoherence in the open system must be accompanied by the irreversible perturbation
of the environment's state but the energy of the environment must be asymptotically preserved.
However, the linear coupling to the bosonic environment implies that the only
change of its state is caused by irreversible processes of emission and absorption
of single bosons which must alter the environment's energy. This statement apparently
contradicts  various models of pure decoherence existing in the literature [7,8].  
\par
In order to explain this discrepancy and to illustrate at the same time the phenomenon
of false decoherence we discuss the simplest
version of the {\it spin - boson model}. The spin-$1/2$ represents an open system 
while bosonic field at the vacuum state a model of a (zero-temperature) bath. 
Although the example
with the Hamiltonian defined below is one of the simplest and most studied exactly solvable
models in quantum theory [9], it seems that some of its subtle mathematical and physical
features were overlooked in the context of decoherence in quantum open systems.
\par
The bosonic reservoir is defined in terms of fields $a(\omega), a^+(\omega)$
satisfying CCR
$$
[a(\omega), a^+(\omega')]= \delta (\omega - \omega')\ ,\ \omega,\omega' 
\in [0 ,\infty)
\eqno(1)
$$
a single-boson Hilbert space $L^2[0,\infty)$, a single-boson Hamiltonian $h_1$,
$(h_1f)(\omega)= \omega f(\omega)$ and the second quantization Hamiltonian
$$
H_B =  \int_0^{\infty} d\omega\,\omega\,a^+(\omega)a(\omega)
\eqno(2)
$$
acting on the bosonic Fock space ${\cal F}_B \bigl(L^2[0,\infty)\bigr)$ with the vacuum
state $\Omega$. Introducing  "smeared fields" $a(f) = \int_0^{\infty}d\omega\, 
{\bar f}(\omega) a(\omega)$ we can define a spin-boson Hamiltonian depending on the 
function ("formfactor") $g\in L^2[0,\infty)$
$$
{\bf H}_g = \sigma_3\otimes \bigl(a(h_1g)+ a^+(h_1 g)\bigr) +\sigma_0\otimes H_B 
\eqno(3)
$$ 
acting on the Hilbert space 
$$
{\cal H}_{SB} = {\bf C}^2\otimes {\cal F}_B \bigl(L^2[0,\infty)\bigr)
\equiv {\cal F}_B\bigl(L^2[0,\infty)\bigr)\oplus {\cal F}_B\bigl(L^2[0,\infty)\bigr)
\eqno(4)
$$
where $\sigma_{\mu} , \mu = 0,1,2,3$ are standard Pauli matrices.
\par
This particular choice of the parametrization of the interaction Hamiltonian becomes 
clear soon. The condition $\|g\|^2 = \int_0^{\infty}|g(\omega)|^2 d\omega <\infty$ 
is usually satisfied by introducing an ultraviolet
cut-off at $\omega_c$ and puting in the infrared region 
$$
|g(\omega)|^2 \sim \omega^{-1+\kappa},\ {\rm with }\ \kappa >0\ .
\eqno(5)
$$
The Hamiltonian (3) can be diagonalized using unitary Weyl operators $W(f)=\exp\{a(f)-a^+(f)\}
$ with $f\in L^2[0,\infty)$
acting on the Fock space ${\cal F}_B\bigl(L^2[0,\infty)\bigr)$ and satisfying
$$
W(f)^* = W(-f),\ W(f)W(h) = e^{-i{\rm Im}<f,h>} W(f+h) 
,\ W(f) a(\omega)W(f)^* = a(\omega) + f(\omega){\bf 1} \ .
\eqno(6)
$$
The vectors $W(f)\Omega$ are called {\it coherent states} and they form an overcomplete set in
the following sense. If for a given vector $\Psi$ from the bosonic Fock space and any 
$f\in L^2[0,\infty)$ $<\Psi , W(f)\Omega> = 0$, then $\Psi =0$. Taking into
account the formula (6) and that $<\Omega ,W(f)\Omega> = \exp\{-(1/2)\|f\|^2\}$ we
obtain
$$
|<W(f)\Omega ,W(g)\Omega>|^2 = e^{-\|f-g\|^2} \leq e^{-|\|g\|-\|f\||^2}\ .
\eqno(7)
$$
As a consequence for $\|g\|=\infty$ $W(g)$ cannot be defined as a unitary operator on 
${\cal F}_B\bigl(L^2[0,\infty)\bigr)$.
Introducing a unitary operator on ${\cal H}_{SB}$ (4) ($\|g\|<\infty$)
$$
{\bf W}(g) =\pmatrix{W(g) & 0      \cr
                     0  & W(-g) \cr}\ .
\eqno(8)
$$
we obtain the diagonalized form
$$
{\bf W}(g) {\bf H}_g {\bf W}(g)^* = \sigma_0\otimes\int_0^{\infty} 
d\omega\,\omega\,a^+(\omega)a(\omega) -  E_g
{\bf 1}
\eqno(9)
$$
where
$$
E_g = <g,h_1 g> = \int_0^{\infty} \omega |g(\omega)|^2 d\omega\ .
\eqno(10)
$$
Therefore the degenerated ground states of ${\bf H}_g$ are given by
$$
{\bf H}_g \Phi_{\pm}(g) = -E_g \Phi_{\pm}(g)\ ,\ \ 
\Phi_{\pm}(g) = e_{\pm}\otimes W(\pm g)\Omega\ ,{\rm where}\  \sigma_3e_{\pm} = \pm e_{\pm}\ . 
\eqno(11)
$$ 
In the standard dynamical approach to decoherence one starts with an initial product state of
the spin-boson system
$$
\Psi_{in} = \psi\otimes\Omega\ ,\ \psi = \psi_-e_- +\psi_+e_+\ ,\  \psi_{\pm} \in {\bf C}
\eqno(12)
$$
satisfying
$$
|<\Psi_{in}|\Phi_{\pm}(g)>|^2 = e^{-\|g\|^2}\ ,\ E(\Psi_{in}) = <\Psi_{in}|{\bf H}_g|\Psi_{in}>
=0\ ,
\eqno(13)
$$
computes its time evolution governed by the Hamiltonian (3)
$$
\Psi(t) = e^{-it{\bf H}_g}\Psi_{in}= 
\exp\{i(tE_g - {\rm Im} <g|g_t>)\} \bigl(\psi_-e_-\otimes W(g_t-g)\Omega +
\psi_+e_+\otimes W(g-g_t)\Omega\bigr)
\eqno(14)
$$
where $g_t(\omega)= e^{-i\omega t}g(\omega)$
and calculates the reduced density matrix for the spin 
$$
\rho_t = {\rm Tr}_B |\Psi(t)><\Psi(t)| =
\pmatrix{|\psi_+|^2 & \psi_+ \overline{\psi}_- e^{-\gamma_t}      \cr
          \overline{\psi}_+ \psi_- e^{-\gamma_t} & |\psi_-|^2  \cr}
\eqno(15)
$$
with
$$
\gamma_t = 2\|g-g_t\|^2\ .
\eqno(16)
$$

The interpretation of the obtained results is rather straightforward. Two degenerated
ground states of the Hamiltonian ${\bf H}_g$ should be interpreted as the states
of a {\it dressed spin} which consists of a {\it bare spin} and a {\it cloud} of virtual bosons
represented by the coherent states $W(\pm g)\Omega$. As for $t\to\infty$ the traveling wave
$g_t$ becomes orthogonal to $g$ the asymptotic form of $\Psi(t)$ possesses the structure
of superposition of two triple product states $e_{\pm}\otimes W(\pm g)\Omega\otimes 
W(\mp g_t)\Omega$.
Therefore, the evolution of the initial product
state (12) given by (14) describes the process of formation of the cloud accompanied by 
emission of the average energy $E_g$ in a form of coherent traveling waves $\pm g_t$. 
In principle, such process may be observed
for example after rapid injection of an electron into a polar medium when the new stable physical 
system - a {\it polaron} - is formed. However, this process has nothing to do 
with the decoherence of polaron states in a solid. Similarly, for fundamental interactions 
(e.g. electromagnetic one) the processes of dressing could be
important in the presence of particle creation or in cosmological context
but not for the low energy decoherence phenomena.
Therefore, from the physical point of view the discussed model describes a phenomenon
which should be called {\it false decoherence} [13].
\par
The degree of false decoherence is characterised by $\gamma_t\leq 8\|g\|^2$ .
To obtain an asymptotically exponential decay of the off-diagonal elements of the
reduced density matrix (15) we must assume that $\|g\| =\infty$ (see (16)) and
moreover
$$ 
0 < \gamma = \lim_{t\to\infty} {\gamma_t\over t} =
\lim_{t\to\infty} \int_0^{\omega_c}\omega^2|g(\omega)|^2 {1-\cos \omega t\over t\omega^2}
\ d\omega = \pi\lim_{\omega\to 0} \omega^2 |g(\omega)|^2\ .
\eqno(17)
$$
This result agrees with a standard wisdom relating the pure decoherence rate to the value
at $\omega = 0$ of the {\it spectral density function} 
$$
{\hat R}(\omega) =  \int_{-\infty}^{\infty}
e^{i\omega t}< R(t)R>_B dt
\eqno(18)
$$
where $R$ is a bath's operator appearing in the interaction Hamiltonian $\sigma_3\otimes R$
and $<\cdot>_B$ is an average with respect to the environment's state. 
It is a special case of the quantum {\it fluctuation - dissipation
theorem} which in fact should be called in this context a "fluctuation-decoherence theorem".
For our model $R = a(h_1 g) + a^+(h_1 g)$ and hence ${\hat R}_0(\omega) = 2\pi\omega^2 
|g(\omega)|^2$ where the subscript "$0$" indicates the zero-temperature (vacuum) state
of the bath.
However, a non zero value of $\gamma$ means
that $|g(\omega)|^2\sim \omega ^{-2}$ what due to (5) corresponds indeed to $\|g\| =\infty$ 
and the average energy $E_g$ (10) emitted during the false decoherence process is infinite.
\par
The meaning of this result is the following. From the previous discussion we know that
only under the
condition $\|g\| <\infty$ the formal expression (3)
defines a Hamiltonian possessing a (doubly degenerated) ground state and describes
a physically admissible stable system. For $\|g\|=\infty$ even its renormalised version
${\bf H'}_g = {\bf H}_g + E_g{\bf 1}$ does not possess a ground state in the Hilbert
space (4). This is a well known {\it van Hove phenomenon} related to the existence of
{\it nonequivalent representations of CCR} and being the simplest instance of difficulties
with the formulation of  a mathematically sound Quantum Field Theory [10].
\par
In principle, the states $\Phi_{\pm} (g) $ treated as limits of "normal" states from the Hilbert 
space (4) with $\|g\| \to\infty$ can exist in the sense of state functionals on the algebra 
of observables. But in this case they are {\it disjoint}, i.e. they define
nonequivalent representations of the algebra of observables. Formally, it follows from the
formula $\Phi_+(g) = \sigma_1\otimes W(2g)\Phi_-(g)$ which for $\|g\|\to\infty$ indicates that
there exists no unitary operator which transforms $\Phi_-(g)$ into $\Phi_+(g)$. 
Physically, it means that
their superpositions are indistinguishable from their mixtures . In other words
{\it superselection rules} appear in the theory and the corresponding {\it classical
observables} emerge. For our model $z$-component of a dressed spin becomes such an observable. 
Some authors 
invoke this mechanism to describe the emergence of classical observables for quantum systems 
and call this phenomenon {\it decoherence} [11]. In fact, it should be called {\it static
decoherence} because the disjointness is a permanent feature of these states.
Although from the mathematical point of view 
this is an atractive approach, on the other hand it can lead to profound interpretational 
difficulties. Strict application of this idea, for example in the case of
electromagnetic interactions, produces a physically nonacceptable superselection rule 
which
prevents coherent superpositions of different momentum states of an electron [10,11].
As the author of [10] writes : "The seeming paradox may serve as a warning against overrating
the significance of idealizations in the mathematical description of a physical situation".
One can hope that the recent rigorous investigations of the nonrelativistic electrodynamics [12]
will improve our understanding of this fundamental issue.
\par
A very similar model of harmonic oscillator heat baths (temperature $T>0$) with a linear 
coupling to an open system is used to model quantum Brownian
motion. In the Markovian approximation the following Master Equation for the reduced density
matrix of the 1-dimensional Brownian particle, the so-called Caldeira-Leggett equation,
has been derived [8]  
$$
{d\over dt}\rho_t = -i[H ,\rho_t] -i\eta [X, \{P,\rho_t\}] -2M\eta T [X,[X,\rho_t]]
\eqno(19)
$$
where $X, P$ are position and momentum operators, $H = P^2/2M +V(X)$, $\eta$ is a 
{\it friction 
constant}, $M$ is a mass of the Brownian particle and V(X) is a potential energy.
While in a general case the eq.(19) describes both decoherence and dissipation the formal
heavy particle limit $(M\to\infty , M\eta = {\rm const.})$ produces a simplified equation describing pure
decoherence
$$
{d\over dt}\rho_t = -i[V(X) ,\rho_t] - \gamma[X,[X,\rho_t]]
\eqno(20)
$$
with the decoherence rate $\gamma >0$. Again we have an apparent contradiction to our 
no-go theorem and the solution of this paradox is also similar. The derivation of
eq.(19) is based again on the condition $\lim_{\omega\to 0}{\hat R}(\omega) >0$.
It is easy to check that for $\omega << T$ (we put $\hbar\equiv k_B \equiv 1$) 
${\hat R}_T(\omega) \simeq (T/\omega){\hat R}_0(\omega)$ and hence the condition of
above is satisfied for the {\it ohmic form} [8,9] of the interaction. In our notation it means 
that $|g(\omega)|^2 \sim \omega ^{-1}$ in the infrared region
and according to (5) it corrresponds to a singular and unstable model.
\par
Summarizing the above results one should stress that the very idea of the theory of 
open systems
demands a clear operational decomposition into the well-defined open system $S$ and the stable 
reservoir $R$. It is possible only if we use an effective theory for "dressed" systems  
with Hamiltonians
of both systems $H_S$ and $H_R$ acting as self-adjoint operators on the Hilbert 
spaces ${\cal H}_S , {\cal H}_R $
of physical ("dressed") states and the interaction $H_{int}$ between $S$ and $R$ is 
a weak perturbation of $H_S + H_R$ with all necessary cut-offs and formfactors such
that $H_S + H_R + H_{int}$ is a well-defined 
Hamiltonian on the Hilbert space 
${\cal H}_S \otimes {\cal H}_R$ possessing a ground state. This last assumption is a minimal
stability condition which must be satisfied even if we consider thermal states of the 
environment only. 
In order to fulfil these requirements for the models of
open systems linearly coupled to harmonic oscillator (bosonic) baths one has to introduce 
beside the ultraviolet cut-off $\omega_c$ the proper scaling of formfactors (5) in the 
infrared domain. It follows that such models are very useful to describe dissipation 
acompanied by decoherence like spontaneous emission of light by atoms and molecules
but fail in the case of pure decoherence, because $\lim_{\omega\to 0} {\hat R}(\omega) =0$. 
As a consequence the models based on vacuum
fluctuations of the background quantum fields (gravitational, electromagnetic,...) [14] 
are very
unlikely to solve the problem of transition from the quantum to classical world.
\par
Fortunately, it is not difficult to construct  proper models of pure decoherence.
They should involve interactions enabling {\it elastic
scattering} processes which perturb the reservoir's state without
changing its energy. Beside the direct elastic collisions with atoms,
molecules, photons etc. the other "bilinear" processes are possible within this scheme 
like, for instance, absorption of a foton followed by an excitation of internal degrees of
freedom of a Brownian particle and the time-reversed process [15]. The simplest 
"spin-boson" version of such model is given by the following Hamiltonian
$$
{\bf H}^f = \sigma_3\otimes \bigl(a^+(f;+)a(f;-)+ a^+(f;-)a(f;+)\bigr) +
\sigma_0\otimes \sum_{\epsilon =+,-}H_B(\epsilon) 
\eqno(21)
$$ 
where
$$
H_B (\epsilon) =  \int_0^{\infty} d\omega\,\omega\,a^+(\omega ;\epsilon)a(\omega ;\epsilon)\ .
\eqno(22)
$$
The bosonic fields satisfy CCR
$$
[a(\omega ;\epsilon), a^+(\omega' ;\epsilon')]= \delta_{\epsilon\epsilon'}
 \delta (\omega - \omega')\ .
\eqno(23)
$$
The single-boson Hilbert space $L^2[0,\infty)\otimes {\bf C}^2$ contains an additional discrete
quantum number $\epsilon = \pm$ and we put a single-boson Hamilonian $h_1$,
$(h_1 f)(\omega ;\epsilon)= \omega f(\omega ;\epsilon)$. The boson can be seen, for instance, 
as a particle moving in 1-dimensional space with the kinetic energy
$\omega$ and the momentum direction $\epsilon$. The smeared fields are obviously
defined by $a(f ;\epsilon) = \int_0^{\infty}d\omega\, {\bar f}(\omega) a(\omega ;\epsilon)$. 
For $\int_0^{\infty}|f(\omega)|^2 d\omega <\infty$ the Hamiltonian (21) is
well-defined on the Hilbert space
$$
{\cal H}^{SB} = {\bf C}^2\otimes {\cal F}_B \bigl(L^2[0,\infty)\otimes {\bf C}^2\bigr)
\equiv {\bf C}^2 \otimes{\cal F}_B\bigl(L^2[0,\infty)\bigr)
\otimes {\cal F}_B\bigl(L^2[0,\infty)\bigr)
\eqno(24)
$$
and possesses a double degenerated product ground state $e_{\pm}\otimes\Omega_+\otimes
\Omega_-$. Instead of the vacuum state we take as a reference state of the environment
a quasi-free state with the average $<\cdot >_B $ describing a free bosonic gas in the 
thermodynamic limit determined
uniquely by the density $n(\omega)$ such that
$$
< a^+(\omega ;\epsilon)a(\omega' ;\epsilon')>_B = n(\omega)\delta_{\epsilon\epsilon'}
\delta (\omega-\omega')\ .
\eqno(25)
$$
We do not go into the detailed analysis of this simple model but we present
only the formula for the pure decoherence rate in the Markovian low density Born 
approximation [16]
$$
\gamma \simeq \pi\int_0^{\infty} |f(\omega)|^4 n(\omega) d\omega\ .
\eqno(26)
$$
From the formula (26) it follows that choosing a proper $f$ one can always reproduce 
a given finite
value of the decoherence rate with arbitrarily small $\|f\|$. No problems with
infinite clouds of virtual bosons or instabilities of the equilibrium state 
appear in this model. As shown in [17] scattering mechanisms can perfectly describe
quantum Brownian motion of a heavy particle immersed in a medium.
\par
The final topic of this Letter is the influence of chaotic properties of an environment
on the decoherence rate in an open system. The intuition supported by some heuristic
arguments suggests, as it is formulated in [18], that "...one would expect that 
environments with
unstable dynamics will be much more efficient decoherers,...". The closer look at this problem 
shows that the oposite statement is true. Namely, for pure decoherence we need an 
irreversible
perturbation of the environment's state which asymptotically conserves its energy. Therefore, 
the reservoir's energy eigenstates should be degenerated and labeled by other quantum 
numbers which
can be altered without energy modification. However, for a chaotic system its energy levels
are typically nondegenerated due to the mechanism of {\it level repulsion} [19]. As a consequence
chaotic reservoirs are worse decoherers. 
\par
The physical arguments of above can be illustrated by the following model introduced
in the context of "$1/f$ noise" in [20]. Consider again a $1/2$-spin system interacting with an 
ensemble of $N$ identical $M$-level chaotic quantum systems by means of the following mean-field type 
Hamiltonian
$$
{\bf H}_Q = \sigma_3\otimes N^{-1/2}\sum _{k=1}^N Q^{(k)} + 
\sigma_0\otimes\sum_{k=1}^N h^{(k)}
\eqno(27)
$$
where $h^{(k)}$ is a copy of the Hamiltonian with the spectral resolution
$h=\sum_{m=1}^M \epsilon_m |m><m|$, $\epsilon_{m+1} \geq \epsilon_m$ 
and $Q^{(k)}$ is a copy of an operator $Q=Q^*$, ${\rm Tr}Q =0$. The rerefence
state of the environment is assumed to be a product state $\otimes_{k=1}^N \rho^{(k)}$
where $\rho^{(k)}$ is a copy of a microcanonical state giving an uniform probability 
distribution over all states $|m>$. Under the above assumption for $N\to\infty$
the mean-field reservoir's observable $N^{-1/2}\sum_{k=1}^N Q^{(k)}$ behaves like a
Gaussian noise and in the Markovian approximation the pure decoherence rate $\gamma$ 
for the spin is given by the following version of the fluctuation-dissipation formula
$$
\gamma = {1\over 2}\lim_{\omega\to 0}{\hat R}(\omega)\ ,\ {\hat R}(\omega) =
{\pi\over M}\sum_{m,m'=1}^M |<m|Q|m'>|^2 \delta \bigl((\epsilon_m - 
\epsilon_{m'})-\omega\bigr)\ .
\eqno(28)
$$ 
The formula makes sense also when instead of identical subsystems the reservoir consists 
of a large random ensemble of chaotic systems with Hamiltonians $h^{(k)}$ characterized by a certain 
average nearest-neighbour level spacing $\Delta$. Then the Wigner level fluctuation law [19]
is applicable and gives the following nearest-neighbour level spacing distribution 
$$
p(s) = (\pi s/2\Delta) \exp (-\pi s^2/4\Delta^2)\ .
\eqno(29)
$$
For $\omega << \Delta$ only the nearest-neighbour level spacings $\epsilon_{m+1} - \epsilon_m$
contribute to the spectral function ${\hat R}(\omega)$ (28). Assuming that  the matrix
elements $<m+1|Q|m>$ are not strongly correlated with $\epsilon_{m+1}-\epsilon_m$
we obtain
$$
{\hat R}(\omega)\simeq \pi {\bar Q}^2 p(\omega)\sim \omega
\eqno(30)
$$
where ${\bar Q}^2$ is an averaged value of $|<m+1|Q|m>|^2$. As a consequence of (28)(30)
pure decoherence rate is equal to zero for chaotic systems while for  regular ones Poisson
distribution of the level spacing gives a finite value of $\gamma$. 
\par

\acknowledgments
The author thanks Micha\l\ Horodecki  for discussions.

\end{document}